\documentclass[epsfig,12pt]{article}
\usepackage{epsfig}
\usepackage{graphicx}
\usepackage{array}
\usepackage{color}
\usepackage{bm}
\usepackage{bm}
\usepackage[nottoc,notlof,notlot]{tocbibind} 
\usepackage[titles]{tocloft}

\newcommand{\beq}{\begin{equation}}   
\newcommand{\eeq}{\end{equation}}
\newcommand{\beqn}{\begin{eqnarray}}   
\newcommand{\eeqn}{\end{eqnarray}}

\newcommand{\gsim}{\lower.7ex\hbox{$
\;\stackrel{\textstyle>}{\sim}\;$}}
\newcommand{\lsim}{\lower.7ex\hbox{$
\;\stackrel{\textstyle<}{\sim}\;$}}
\usepackage{showlabels}

\begin{document}

\setcounter{footnote}{0}

\begin{center}

{\small HISTORICAL CURIOSITY: \\
HOW ASYMPTOTIC FREEDOM OF THE YANG-MILLS
THEORY COULD HAVE BEEN DISCOVERED 
THREE TIMES BEFORE GROSS, WILCZEK, AND POLITZER, BUT WAS NOT\,\footnote{This article was published in {\bf 2001}, see Festschrift  \cite{Ioffe}. I added a couple of sentences in 2022.} }

\vspace{3mm} 

{\large \bf M. Shifman}

\end{center}

\begin{center}

{\em William I. Fine Theoretical Physics Institute, University of Minnesota,
Minneapolis,
MN 55455, USA}

\vspace{3mm}

{\bf Abstract}

\end{center}

{\small\it Asymptotic freedom as the basic property of QCD was discovered
by Gross, Wilczek, and Politzer in 1973. Personal recollections of David 
Gross which are being published in this Volume
vividly describe the historical background and the chain of events
which led to this fundamental breakthrough. Unfortunately, I failed  to 
obtain  Politzer's side of the story. Some details can be 
found in an interview which Prof. Politzer gave  
to R. Crease and C. Mann on February 21, 1985 \cite{DP}.  Below I acquaint the
reader with the pre-1973 appearances of asymptotic freedom which,
unfortunately, went unnoticed. }

\vspace{4mm}

\setcounter{equation}{0}

The {\em anti}screening of the gauge coupling 
constant in  the non-Abelian gauge
theories had been detected  three times before the actual discovery of 
asymptotic freedom  was made by David Gross, Frank Wilczek and David Politzer. In two 
instances, the
 results of calculations exhibiting the antiscreening were
reported in the journal publications \cite{Tere,Khri} in the 1960's
(V.~Vanyashin and M. Terentev, 1965; I. Khriplovich, 1969).
The third time this calculation was apparently  carried out in 1971
by G. 't~Hooft; the result was not published, however.
Since this story is rather curious (and, perhaps, even instructive, in 
retrospect), I think, the reader deserves to be acquainted with it.
I hasten to add, though, that the above authors did not recognize
the importance of the antiscreening for the theory of strong interactions;
nor did they grasp possible implications. Therefore, by no means I 
want to imply that the discovery of asymptotic freedom  was made 
before 1973. There can be no questions -- the true discoverers of
asymptotic freedom are  Gross, Wilczek and Politzer. 

The paper of Vanyashin and Terentev,  titled 
{\em The Vacuum Polarization of 
a Charged Vector Field}, was published\,\cite{Tere} in 1965. 
The authors considered propagation of the charged massive vector field 
in the background electromagnetic field. They obtained the 
 Heisenberg--Euler  effective Lagrangian at  one loop. 
The calculation was carried out  in the unitary gauge.
In modern terms
one can say that the authors found the one-loop
effective Lagrangian in the SU(2) massive Yang-Mills theory.

Although the mass of the vector field was treated as hard, rather than
in the framework of the Higgs mechanism, in the
one-loop approximation the theory is self-consistent.
The power divergences cancel. The only remaining divergence is 
logarithmic; it is exactly the same as in the SU(2) Yang-Mills theory
with the Higgs mechanism. 
The coefficient in front of 
\begin{equation}
\frac{e^2}{32\pi^2} \, E^2\,  \ln\frac{M_{\rm UV}^2}{eE}
\label{VT}
\end{equation}
in the large $E$ limit
is nothing but the first coefficient in the $\beta$ function.
(Here $M_{\rm UV}$ is the ultraviolet cutoff used in
calculating the Heisenberg--Euler Lagrangian).
Vanyashin and Terentev found this coefficient to be equal to
\begin{equation}
-7 = - \left( \frac{22}{3}-\frac{1}{3}\right)\,,
\end{equation}
(see Eq. (30) in their work). The first term,
$-{22/}{3}$, is what one would get for the massless Yang-Mills field,
while $1/3$ is the contribution of the longitudinal $W$ bosons.
In the theory with the Higgs mechanism of the mass generation
this $1/3$ emerges as the contribution of the Higgs field.

Vanyashin and Terentev perfectly realized that 
the coefficient  they found
represented the charge renormalization.
Let me quote a passage from the concluding 
part 
of the paper: ``Unlike conventional electrodynamics, where the 
renormalized 
charge is smaller than the bare one, our case gives $e^2>e_0^2$ [...] 
As is 
well-known, in electrodynamics the restriction $0\leq Z\leq 1$ on 
the $Z$ 
factor (the charge renormalization) follows from the 
most general principles [...]. In our case it is impossible, generally 
speaking,
to get the restriction $0\leq Z\leq 1$ starting from these 
principles [...].
Nevertheless, the inequality $Z>1$ seems extremely undesirable."\footnote{In 2008 Arkady Vainshtein wrote (private communication to G. t'Hooft):
``Valentin Zakharov who was at Terentev's seminar recollects [...] that
the main problem Isaac Pomeranchuk had with Terentev's work was positivity of the spectral function in the K\"allen-Lehman representation of the polarization operator. Pomeranchuk, of course, knew very well that $Z>1$ implies asymptotic freedom in the UV but he viewed the theory of massive charged vector field as nonphysical, particularly, because of the positivity problem.  Terentev was not able to give him a satisfactory answer.
[...]  A simple relation between the polarization operator  and the coupling constant will be subsequently lost.  A derivation based on scattering amplitudes was carried out 
in  \cite{1}  
  and later by 
L. Lipatov when he considered reggezation."} 

For a short time, around 1970, Terentev was my thesis adviser, and I
had the opportunity of discussing various physics issues with him.
Certainly, he was expert in the renormalization group equations,
and was well aware of  the fact that $ Z\leq 1$ in QED implied the
growth of
the effective charge in QED at short distances, and {\em vice versa}.
In the ITEP theory department this was a ``$2\times 2 = 4$" statement. 
Landau, Abrikosov and  Khalatnikov at first made a mistake in their 
1954
work\,\cite{LAK} on the effective charge in QED, and got 
the $ \beta$ function negative. For some time
Landau was entertaining himself working out the implications
of the negative $ \beta$ function -- the fall off of the 
effective charge at short distances, and so on. 
He recognized the importance of the phenomenon right away and widely
discussed it with his ITEP colleagues and elsewhere.
Only somewhat later a sign error was found
by Boris Ioffe and Alexei Galanin (See Geshkenbein's note).
In the published version\,\cite{LAK} the zero charge result 
(i.e. infrared freedom) was reported.

Why Terentev never returned to his 1965 work after the results of the
SLAC-MIT experiments on deep inelastic scattering
became known, remains a mystery. The experiments on deep inelastic
scattering and the subsequent works of Bjorken and Feynman
were subject to intense scrutiny
in ITEP theory division, prior to the discovery of asymptotic freedom
in 1973. 

Khriplovich's work\,\cite{Khri} 
was specifically devoted to the one-loop corrections in (massless)
non-Abelian Yang-Mills theories. A non-covariant 
(Coulomb)  gauge was used, which is ghost-free. In this gauge the charge 
renormalization is given by that of the $D_{00}$ component of the 
gauge field Green 
function, much in the same way as in QED. The logarithmically divergent 
part of $D_{00}$ is presented  in 
Eq. (61) of Ref. \cite{Khri}, from which it is straightforward to read
off the corresponding $Z$ factor (in the first line see the first two terms in the square brackets). After
taking into account an  unconventional 
normalization of the coupling constant there,  one  immediately 
recovers the first coefficient in the $\beta$ function,  the famous 
$-22/3$ (the theory considered in Ref. \cite{Khri}
was the SU(2) gauge theory). 

Moreover, one  additionally learns  from the discussion around
Eq. (61) in \cite{Khri}  that 
$-22/3$ 
is actually $-8+\frac{2}{3}$. The logarithms are of two different types. 
One  is non-dispersive, 
it corresponds to the instantaneous Coulomb interaction 
and  is responsible for  antiscreening (``minus eight").\footnote{For more
details see Sec. 1.3  of Ref. \cite{nov}.}
  The second logarithm is normal 
dispersive; in full accordance with the general principles it gives
a  positive contribution (``two thirds") characteristic of the normal
screening.  The formula $-\frac{22}{3} = 
-8+\frac{2}{3}$ and the corresponding interpretation of 
antiscreening {\em vs.} screening was rediscovered in
the West only  eight years later \cite{Appelquist:1977tw}.\footnote{Khriplovich was far away, at the Novosibirsk Institute for Nuclear Physics,
and, according to Vainshtein, was less knowledgeable  about implications of his  $Z>1$ result for
the $\beta$ function. Alas... he missed the connection.}

\vspace{0.2cm}

Now I come to 't~Hooft's story. 
Here is an excerpt from  1998 't~Hooft's talk\,\cite{hoo} where 't~Hooft
recollects on his work in 1971.

\vspace{0.1cm}

``To many physicists, the Bjorken scaling appeared to be a signal that 
no quantum field theory would be fit to account for the observed strong
interactions. Why then did nobody compute the $\beta$  function for the
Yang-Mills theories? I {\em had} computed this function, and knew 
about its
sign. In 1971, as soon as we knew how to renormalize the theory, I 
started a
preliminary calculation, showing the negative sign. Not realizing that this 
was
something totally novel, I could not imagine that this was all that was 
needed
to resolve the outstanding problems of strong interaction theory. I 
actually
thought that this scaling behavior was also known, but not considered
relevant for partons. Since I could not understand their problems, I 
turned
away from the strong interactions that had to be infinitely complicated.

By 1972, when Veltman and I had worked out many examples of gauge
models, we knew how to compute the renormalization counter terms, 
and I had found a way to relate these to the equations governing the 
running
coupling constants. Veltman was not interested in these findings. 
According to
him, they were only relevant for  Euclidean space, which has nothing to 
do
with experimental observations, all done in Minkowski space. On several
occasions I mentioned to him the possibility of a pure gauge theory for 
the
strong interactions, but his mind was made up. As long as they produced 
no
answer to the quark confinement problem, such ideas are useless, and 
no
referee would accept a publication suggesting such an idea. 

In June 1972, a small meeting was organized in Marseilles\,\footnote{{\sl Colloquium on Renormalization of 
Yang-Mills Fields and Applications to Particle Physics}, Centre de Physique Th\'eorique, June 19, 1972, Marseilles, France.}
by Christiaan P.~Korthals~Altes. Symanzik was there. Upon arrival, right at the 
airport, Symanzik 
and I began to discuss our work. Symanzik explained that he worked on 
$\lambda\phi^4$ theory with negative $\lambda$, because of its 
interesting scaling properties. He had been attracted by the idea that 
theories
with a negative $\beta$ function would reproduce parton-like 
behavior, but,
the only models with such properties that he could think of were 
theories
with a negative squared coupling constant, $g^2<0$ or
$\lambda<0$.

Naturally, I asked him how he would solve the problem that the energy
appears to be unbounded from below. He explained that perhaps this
problem could be cured by non-perturbative effects, which at that
time were badly understood. But I informed him of my finding that the
coefficient determining the running of the coupling strength, that he 
called 
$\beta(g^2)$, for gauge theories is negative. Symanzik was surprised and
skeptical. ``If this is true, it will be very important, and you should 
publish
this result quickly, and if you won't, somebody else will,'' he said.

In the meeting, Symanzik ended his talk on negative $\lambda\phi^4$
theories by the remark that further study on the $\beta$ functions
for non-Abelian gauge theories were called for. When it was
 time for questions I came forward to write down on the blackboard the
expression I had derived (with a negative coefficient) [...].

Why did I not follow Symanzik's sensible advice to publish this result?
Weinberg had written a long paper in which he reported about the
complicated infrared structure of gauge theories, which made me 
believe he knew everything about what was to be called infrared 
slavery, 
a  direct consequence of asymptotic freedom. Surely, I thought, he 
understands the scaling behavior of gauge theories. David Gross had 
explained to me that the Bjorken scaling cannot be accommodated for in 
any 
field 
theory, Abelian or not. Veltman was telling me that scaling all 
components 
of the momentum vector would lead you far away from the mass shell,
whereas {\em all} experiments were looking at on-mass-shell hadronic 
final states, so the deep-Euclidean region was unphysical and 
uninteresting."

\vspace{0.5cm}

I am grateful to L. Frankfurt and A. Vainshtein for reminding me 
certain details in the Vanyashin-Terentev work.

\end{document}